\documentstyle[namedreferences,NATO,epsf]{CRCKAPB} 

\def\etal{{et\thinspace al.}\ }

\newcommand{\ion}[2]{\mbox{#1\,{\small #2}}}

\begin{opening}
\title{
FUSE Spectroscopy of the Two Prototype White \protect\\ Dwarfs With Signatures of a Super-hot
Wind}
\author{K. WERNER}
\author{S. DREIZLER} 
\institute{Universit\"at T\"ubingen, Germany}
\author{J.W. KRUK}
\institute{Johns Hopkins University, U.S.A.}
\author{M.L. SITKO}
\institute{University of Cincinnati, U.S.A.}
\end{opening}
\begin{document}

\vspace{-0.4cm}
\noindent
The \ion{O}{VIII} phenomenon describes the occurrence of ultra-high ionization
absorption lines of the CNO elements (e.g.\ \ion{O}{VIII}, \ion{N}{VII},
\ion{C}{VI}, and even \ion{Ne}{X}) in the optical spectra hot of DO WDs. This
requires temperatures of almost $10^6$\,K. Since the first discovery of two
such objects (Werner \etal 1995) we realized that 50\% of all DO WDs are
affected (Dreizler \etal 1995), hence it could be that all DO WDs
experience this phenomenon at the beginning of the WD cooling sequence. The
\ion{He}{II} profiles are symmetric, hence line formation occurs in hydrostatic
layers. However, we cannot uniquely fit the unusually strong profiles with any
model atmosphere, preventing precise parameter determinations. The optical high
ionization CNO lines have pronounced blue wings, corresponding to outflow
velocities of up to several thousand km/s. We stress that none of the objects
was detected in the X-ray region. The inferred low X-ray luminosity confirms
that we do not see absorption lines from super-hot photospheres, but that the
high temperatures are possibly localized along shock fronts in the wind.

\begin{figure}[ht]
\epsfxsize=\textwidth
\epsffile{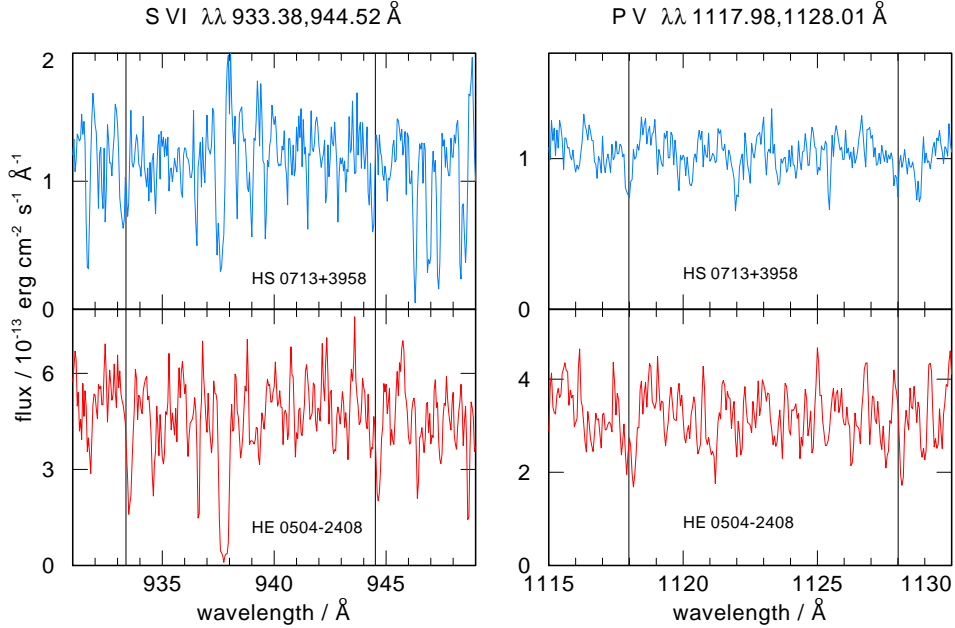}
\caption{Identification of sulfur and phosphorus lines in the FUSE spectra}
\vspace{-0.4cm}
\end{figure}

In the UV range (IUE, HST) no high ionization CNO line is detectable, except
possibly \ion{O}{VIII}~2978\AA\ (Werner \etal 1997, 1999). This negative result
is still not understood. On the other hand we see two unidentified broad
absorption features at 1420\AA\ and 1500\AA. The 1500\AA\ feature may be 
the blue shifted absorption component from a P~Cygni profile of
\ion{C}{IV}~1550\AA. This is corroborated by a very weak and broad emission
bump near the rest wavelength. The terminal wind speed is about 10\,000\,km/s
and mass-loss could be driven by radiation pressure. The only photospheric line
clearly identified is \ion{He}{II} 1640\AA. The presence of \ion{O}{V}~1371\AA\
is uncertain and the \ion{C}{IV} resonance doublet could be of interstellar
origin.

\noindent
We performed FUSE observations of the two prototype DOs. HS0713\-+3958 was
recorded on Nov 21, 2000, for 9825 sec and HE0504-2408 on Dec 5, 2001, for 6625
sec. The spectra confirm earlier ORFEUS observations (Werner \etal 1999) showing
that the FUV spectral shape is surprisingly flat. Strong reddening by
interstellar dust alone is unlikely because no 2200\AA\ feature is seen (E(B-V)
below 0.05). With the superior capabilities of FUSE it was hoped to identify
opacity sources responsible for the flat spectral shape, however, we see no
hint to any potential absorbers, probably due to insufficient S/N. It is
possible, that a dense curtain of weak lines from the iron group elements
remains undetected. Note that a similiar flat UV shape observed in the sdOB
EC11481-2303 results from heavy line blanketing (see Hammer \etal this
proceedings). The usual ISM absorption lines can be identified. H$_2$ is very
weak in HE0504-2408 but prominent in HS0713+3958. The only photospheric lines
that can be identified in both objects are from \ion{S}{VI} and \ion{P}{V}
(Fig.\,1). In addition, HS0713+3958 shows a weak \ion{O}{VI} resonance doublet
which, however, can also stem from the ISM.

\end{document}